\documentclass[a4paper]{iopart}
\pdfoutput=1
\usepackage{graphicx}
\usepackage{enumerate}
\usepackage{iopams}

\usepackage{hyperref}
\usepackage{textcomp}
\usepackage{pgf}

\let\eqref\eref

\def\bra#1{\langle#1 |}
\def\ket#1{| #1\rangle}

\def\D{\mathrm{\scriptscriptstyle D}}
\def\GGE{\mathrm{\scriptscriptstyle GGE}}

\def\dav#1{\langle #1 \rangle_{\D}}
\def\ggeav#1{\langle #1 \rangle_{\GGE}}
\def\tav#1{\overline{#1}}
\def\ud{\mathrm{d}}
\def\ee{\mathrm{e}}
\def\HA{\hat{H}_{\mathrm{A}}}
\def\Hlrh{\hat{H}_{\mathrm{lrh}}}
\def\IPR{\mathrm{IPR}}
\def\op#1{\hat{#1}}
\begin{document}

\title{How to calculate quantum quench distributions with a weighted Wang-Landau Monte Carlo}

\author{Simone Ziraldo$^{1,2}$ and Giuseppe E. Santoro$^{1,2,3}$}
%
\address{$^1$ SISSA, Via Bonomea 265, I-34136 Trieste, Italy}
\address{$^2$ CNR-IOM Democritos National Simulation Center, Via Bonomea 265, I-34136 Trieste, Italy}
\address{$^3$ International Centre for Theoretical Physics (ICTP), P.O.Box 586, I-34014 Trieste, Italy}

\begin{abstract}
We present here an extension of the Wang-Landau Monte Carlo method which allows us to get
very accurate estimates of the full probability distributions of several observables after
a quantum quench for large systems, whenever the relevant matrix elements are calculable, but the
full exponential complexity of the Hilbert space would make an exhaustive enumeration impossible beyond
very limited sizes.
We apply this method to quenches of free-fermion models with disorder, 
further corroborating the fact that a Generalized Gibbs Ensemble
fails to capture the long-time average of many-body operators when disorder is present. 
\end{abstract}


\date{\today}
\maketitle
\tableofcontents 

\section{Introduction} \label{sec:intro}

A sudden quench of the Hamiltonian parameters is perhaps the simplest 
form of out-of-equilibrium dynamics that a closed quantum system can experience.
Experiments with ``virtually isolated'' cold atomic species in optical lattices~\cite{Bloch_RMP08,Lewenstein_AP07} 
have transformed this seemingly theoretical dream into a rich and lively stage. 
Several fundamental issues of theoretical quantum statistical physics, like the onset of thermalization 
which is generally expected to occur for a closed quantum system after a sudden quench \cite{Deutsch_PRA91,Sred_PRE94,Rigol_Nat},
or the ``breakdown of thermalization'' \cite{Rigol_PRL09, Rigol_PRA09} expected when the
system is integrable or nearly integrable, are now of experimental relevance \cite{Kinoshita_Nat06}.
We refer the reader to a recent review \cite{Polkovnikov_RMP11} for an extensive introduction to such non-equilibrium 
quantum dynamics issues. 

The issue we want to tackle in this paper is the following. Suppose you perform a 
quantum quench of the Hamiltonian parameters, abruptly changing, at $t=0$, from $\hat{H}_0 \to \hat{H}$. 
If $|\Psi_0\rangle$ denotes the initial quantum state at $t=0$, and $|\alpha\rangle$ 
the eigenstates of $\hat{H}$ with energy $E_{\alpha}$, the ensuing quantum dynamics 
would lead to averages for any given operator $\op{A}$ given by:
\begin{equation} \label{defO:eqn}
\hspace{-10mm}
A(t) \equiv \langle \Psi_0 | e^{i\op{H}t} \hat{A} e^{-i\op{H}t} | \Psi_0\rangle  
      = \sum_{\alpha} |c_{\alpha}|^2 A_{\alpha\alpha} +
        \sum_{\alpha'\ne \alpha} e^{i(E_{\alpha'}-E_{\alpha})t}
       c_{\alpha'}^* A_{\alpha'\alpha} c_{\alpha}  \;,
\end{equation}
where $c_{\alpha}\equiv \langle \alpha | \Psi_0\rangle$ and $A_{\alpha'\alpha} \equiv \bra{\alpha'} \hat{A} \ket{\alpha}$.
The first (time-independent) term in the previous expression dominates the long-time
average of $A(t)$, and is usually known as {\it diagonal average} \cite{Rigol_Nat}
\begin{equation}
\dav{\hat{A}} \equiv \sum_{\alpha} \left| c_\alpha \right|^2 A_{\alpha \alpha} \;.
\end{equation}
%
To calculate it, in principle, we should take the sum over all the (many-body) eigenstates $\ket{\alpha}$
of $\hat{H}$ --- an exponentially large number of states ---, calculating for each of them the overlap
$c_\alpha$ and the associated diagonal matrix element $A_{\alpha\alpha}$.
Luckily, $\dav{\hat{A}}$ can be calculated for many problems, notably those that
can be reduced to quadratic fermionic problems, by circumventing in one way or
another exponentially large sums: for instance, through a detour to time-dependent
single-particle Green's function and the use of Wick's theorem, 
see e.g. ~\cite{art:Ziraldo_PRL_2012,art:Ziraldo_PRB_2013}.
%
%
But suppose that you want to know more than just the diagonal average $\dav{\hat{A}}$,
and pretend to have information on the whole distribution of the values of $A_{\alpha \alpha}$
accessed after the quench~\cite{art:Rigol_Nat_2008,art:BiroliPRL}, i.e., 
\begin{equation}\label{eq:rhod}
\rho_{\D}(A) \equiv \sum_{\alpha} \left| c_\alpha \right|^2 \delta(A-A_{\alpha \alpha}) \;,
\end{equation}
of which $\dav{\hat{A}}$ is just the average: $\dav{\hat{A}}=\int \! \ud A \, \rho_{\D}(A) \; A$. 
Here there is, evidently, a problem: knowing the distribution of $A$ requires exploring
the full many-body Hilbert space, summing over the eigenstates $\ket{\alpha}$,
and this exhaustive enumeration would restrict our calculations to exceedingly small sample sizes, 
although all information on $c_\alpha$ and $A_{\alpha\alpha}$ might in principle be easy to
calculate, or in any case accessible, for instance by just solving a one-body problem (hence, for much larger sizes).
A similar problem occurs in considering, for instance, the corresponding~\footnote{The generalized 
Gibbs ensemble is the relevant ensemble for quenches with quadratic fermionic models, but a similar situation occurs 
for the usual statistical ensembles, i.e., microcanonical, canonical and grand-canonical.}
generalized Gibbs ensemble (GGE) average~\cite{Rigol_PRB06,Rigol_PRL07,Barthel2010,Calabrese2011,Cazalilla2012}
\begin{equation}\label{eq:WL-GGE-av}
\ggeav{\hat{A}} \equiv \sum_{\alpha} \frac{\ee^{-\sum_\mu \lambda_\mu I_\mu^\alpha}}{Z_{\GGE}} A_{\alpha \alpha} \;,
\end{equation}
where $\lambda_\mu$ are Lagrange multipliers which constrain the mean value of each of the constants of
motion $\op{I}_\mu$ to their $t=0$ value,
$\bra{\Psi_0} \op{I}_\mu \ket{\Psi_0} = \Tr\left[ \op{\rho}_{\mathrm{\scriptscriptstyle GGE}} \op{I}_\mu \right]$,
$I_\mu^\alpha \equiv \bra{\alpha} \op{I}_\mu \ket{\alpha}$,  
$Z_{\GGE}$ is the GGE partition function, and 
$\op{\rho}_{\mathrm{\scriptscriptstyle GGE}} \equiv e^{-\sum_\mu \lambda_\mu \op{I}_\mu}/Z_{\GGE}$.
Once again, for ``quadratic problems'' this average is rather simply calculated in terms of single-particle quantities,
but the corresponding distribution
\begin{equation}\label{eq:rhogge}
\rho_{\GGE} (A) \equiv \sum_{\alpha} \frac{\ee^{-\sum_\mu \lambda_\mu I_\mu^\alpha}}{Z_{\GGE}}
\delta ( A - A_{\alpha\alpha}) \;, 
\end{equation}
requires a difficult sum over the Hilbert space.
\footnote{For the GGE ensemble things might be worked out by an appropriate representation of the Dirac's delta, 
or by computing the moment generating function of $\rho_{\GGE}(A)$. 
These tricks however would not work, for instance, for the microcanonical distribution.}

Concerning the issue of thermalization after a quantum quench, we might indeed expect that, if the system is well
described by a GGE ensemble, not only the mean values of $\rho_{\D}(A)$ and $\rho_{\GGE}(A)$ are equal, 
i.e., $\dav{\hat{A}} \equiv \int \ud A \, \rho_\D (A) A = \ggeav{\hat{A}} \equiv \int \ud A \, \rho_\GGE (A) A$,
but also the two distributions should be closely related;
at least this is what a good statistical ensemble should do.

Quite generally, we might formulate the problem as follows: 
how can we obtain information on weighted distributions (or density of states)
\begin{equation} \label{eq:wdist-intro}
\rho_{\mathrm{w}} (A) \equiv \sum_\alpha w_{\alpha} \delta (A - A_\alpha) \;,
\end{equation}
with positive weights $w_{\alpha}$, 
when both $w_{\alpha}$ and $A_{\alpha}$ are  ``easily calculated'', 
but the sum over $\alpha$ runs over an exponentially large ``configuration space''?
As discussed before, examples of this are the diagonal distribution,
where $w_{\alpha}= |c_\alpha|^2$, the GGE distribution, 
where $w_{\alpha}=\ee^{-\sum_\mu \lambda_\mu I_\mu^\alpha}/Z_{\GGE}$, but
also the microcanonical distribution, where $w_{\alpha}$ is a window characteristic function 
for the microcanonical shells, etc.
A Monte Carlo algorithm to perform such exponentially large sums in configuration space
seems unavoidable. We stress that this is so even if one is considering quenches in free-fermion models,
where the relevant many-particle states $|\alpha\rangle$ and matrix elements are easy to write down and calculate.
The alternative of using exact diagonalization methods would put a strong limit on the size of the problem which
can be studied.

In this paper we introduce a Monte Carlo method --- obtained by a rather natural 
extension of the Wang-Landau algorithm (WLA)~\cite{art:WL-PRL,art:WL-PRE,art:WL-CC} ---
which will allow us to compute weighted distributions of the form of Eq.~\eqref{eq:wdist-intro}.
The Wang-Landau algorithm, proposed in 2001 by F. Wang and D.P. Landau, 
is a Monte Carlo method designed to compute the density of states
of a classical statistical mechanics problem. 
The algorithm performs a non-Markovian random walk to build the density of states 
by overcoming the prohibitively long time scales typically encountered near phase transitions 
or at low temperatures.
Besides the classical Ising and Potts models studied in the original 
papers~\cite{art:WL-PRL,art:WL-PRE,art:WL-CC}, the method has been applied to the solution 
of numerical integrals~\cite{art:Belardinelli_PRE_2008}, folding of proteins~\cite{art:Ojeda_BJ_2009}
and many other problems.

Here is the plan of the paper.
In Sec.~\ref{sec:WWLA} we show how the WLA can be extended to compute weighted density of states.
In Sec.~\ref{sec:WL-QuantumQuenches} we show how the weighted-WLA can
be used to compute distributions related to quantum quenches with quadratic fermionic
models. Finally, Sec.~\ref{sec:conclusions} contains a summary and future perspectives.

\section{Weighted Wang-Landau algorithm}\label{sec:WWLA}
%
Let us consider a system with a discrete configuration space, where configurations
can be labeled with an index $\alpha$. 
Given a physical observable $\hat{A}$, we define its weighted (coarse-grained) density of states:
\begin{equation}\label{eq:wdist}
\rho_{\mathrm{w}} (A) \equiv \sum_\alpha w_{\alpha} \delta_{A A_\alpha} \; ,
\end{equation}
with $w_\alpha$ a {\em positive} weight. 
Here $\delta_{A A_{\alpha}}$ is a Kronecker delta, or, if the possible values of $A_{\alpha}$ are too dense to keep them all,
a suitable histogram-window-function coarse-graining of the Dirac's delta.
When $w_\alpha=1$, we recover the usual density of states $\rho(A)$, and the WLA can be used to estimate 
it~\cite{art:WL-PRL,art:WL-PRE,art:WL-CC}.
We will now show that, by properly modifying the WLA, we can compute $\rho_{\mathrm{w}}(A)$ for generic $w_\alpha$s. 

To understand the gist of the approach, consider a generic positive function $\tilde{\rho}_\mathrm{w}(A)$ 
--- which is our best guess for the desired $\rho_{\mathrm{w}}(A)$ ---, and set up a Markov chain random walk 
in which, given a state $\alpha$, a new state $\alpha^\prime$ is generated with a trial probability 
$T(\alpha^\prime | \alpha)$,
which we will take to be symmetric, $T(\alpha^\prime | \alpha)=T(\alpha | \alpha^\prime)$,
and accepted with probability:
\begin{equation}\label{eq:MCstepw}
R(\alpha^\prime | \alpha) = 
\mathrm{Min} \left[1, \frac{w_{\alpha^\prime}}{w_{\alpha}}  
                                  \frac{\tilde{\rho}_\mathrm{w} (A_\alpha)}{\tilde{\rho}_\mathrm{w} (A_{\alpha^\prime})} 
                                  \frac{T(\alpha |\alpha^\prime)}{T(\alpha^\prime | \alpha)} \right] \;.
\end{equation}
With this standard Metropolis Monte Carlo prescription, we know that, after an initial transient, we will
visit the configurations $\alpha$ with an equilibrium distribution $P^{\rm eq}_{\alpha}$ fulfilling the detailed balance condition 
and given by: 
\[ 
P^{\rm eq}_{\alpha} = C \frac{w_\alpha}{\tilde{\rho}_\mathrm{w}(A_\alpha)} \;,
\] 
where $C$ is a normalization constant. 
As in the WLA \cite{art:WL-PRL}, while the random walk goes on, we collect a histogram $h(A)$, 
updating $h(A_\alpha) \to h(A_\alpha) + 1$ at each visited state $\alpha$. 
At equilibrium, after $N_s$ steps, the ``mean'' histogram will then be given by:
\begin{equation}
h(A) = N_s \sum_\alpha P^{\rm eq}_{\alpha} \, \delta_{A A_\alpha} 
= N_s C \, \sum_{\alpha} \frac{w_\alpha}{\tilde{\rho}_\mathrm{w}(A_{\alpha})} \delta_{A A_\alpha} 
= N_s C \, \frac{\rho_\mathrm{w}(A)}{\tilde{\rho}_\mathrm{w}(A)} \;.
\end{equation}
Exactly as for the WLA \cite{art:WL-PRL}, if our guess for $\tilde{\rho}_\mathrm{w}(A)$ 
is a good approximation to $\rho_\mathrm{w}(A)$, the histogram $h(A)$ will be ``almost flat'' (see below).
Obviously, during the random walk, together with the histogram $h(A)$ we also update our guessed 
$\tilde{\rho}_\mathrm{w}(A)$.
Therefore, closely inspired by the WLA \cite{art:WL-PRL}, we propose the following algorithm:
\begin{enumerate}[(1)]
\setcounter{enumi}{-1}
\item Fix a modification factor $f>1$, and set $\ln \tilde{\rho}_\mathrm{w}(A)=0$
and $h(A) = 0$ for all values of $A$;
\item Start the Monte Carlo procedure using Eq.~\eqref{eq:MCstepw}
and update at each step the histogram and the weighted density of states with the rules
$h(A_\alpha) \rightarrow h(A_\alpha) + 1$ and 
$\ln \tilde{\rho}_\mathrm{w} (A_\alpha) \rightarrow \ln \tilde{\rho}_\mathrm{w} (A_\alpha) + \ln f$; 
\item Stop the random walk when $h(A)$ is ``almost flat'' 
(for instance \cite{art:WL-PRL}, when $h(A) > 0.8 \overline{h}$ for all values of $A$, 
where $\overline{h}$ is the mean histogram value).
For the previous observations, at the end of this step $\ln \tilde{\rho}_\mathrm{w} (A)$ is a good approximation 
to $\ln \rho_\mathrm{w} (A)$ with a discrepancy of order $\ln f$;
\item Reduce the value of $f \to \sqrt{f}$, reset $h(A)=0$ and restart the procedure from step (1) 
using the $\tilde{\rho}_\mathrm{w} (A)$ just obtained. 
Stop this loop when $\ln f$ is smaller than the desired discrepancy $\epsilon$.
\end{enumerate}
A similar extension of the WLA has been already been introduced for the particular case in which 
$w_\alpha$ is the Boltzmann distribution, with the aim of computing the free energy profile as a function of a reaction 
coordinate~\cite{art:Kim_JCP_2002,art:Muller_Lect}. 
In the present paper, we will use this algorithm to compute distributions where the weights are not Boltzmann-like, 
but rather associated to quantum quenches.

Let us return for a moment to the original WLA. 
A first trivial observation is that, as it should be, the weighted-WLA with 
$w_\alpha = 1/N$ coincides with the WLA.
In many situations, when the size of the configuration space is too big and the density of states 
ranges over too many orders of magnitude, it is convenient, in computing $\rho(A)$, to run many
WLA over small domains $\Delta^{(i)}_A = [ A_{\mathrm{min}}^{(i)}, A_{\mathrm{max{\phantom i}\!}}^{(i)} ]$. 
But then the update rule of the standard WLA has to be changed to avoid that, during
the random walk, $A_\alpha$ leaves the domain $\Delta^{(i)}_A$.
This trick was already used in the first papers by Wang and Landau, when dealing with the largest 
sizes~\cite{art:WL-PRL}. To avoid leaks from $\Delta^{(i)}_A$, the empirical solution was to reject any 
proposal to states $\alpha^\prime$ with $A_{\alpha^\prime} \notin \Delta^{(i)}_A$, 
without any update of $\tilde{\rho}(A)$ and $h(A)$.
With this prescription, however, there are ``boundary effects'', actually a systematic underestimation 
of the density of states at the borders of the intervals~\cite{art:Schulz_PRE_2003}.
%
Schulz~\textit{et al.}~\cite{art:Schulz_PRE_2003} showed phenomenologically that
such boundary effects are eliminated by using the rather obvious update rule: given a proposal 
$\alpha^\prime$, if $A_{\alpha^\prime}$ is outside the interval we remain in $\alpha$ and 
we update $h(A)$ and $\ln \tilde{\rho}(A)$ using the state $\alpha$, otherwise we accept $\alpha^\prime$
with the usual rule.
This update rule is just what is obtained, rigorously, by using our weighted-WLA.
Indeed, the density of states in a restricted range $\Delta^{(i)}_A$
is proportional to a weighted density of states
in which $w_\alpha = 1$ when $A_\alpha \in \Delta^{(i)}_A$, and zero otherwise.
With these weights, the update rule of our weighted-WLA is exactly the one obtained 
phenomenologically by Schulz~\textit{et al.}~\cite{art:Schulz_PRE_2003}.

\section{Quantum quenches}\label{sec:WL-QuantumQuenches}
%
In this section we come back to the initial problem of computing the distributions 
$\rho_{\D}(A)$ and $\rho_{\GGE}(A)$ related to quantum quenches.
We will show that with the weighted-WLA we can compute these distributions 
for sizes inaccessible with an exhaustive enumeration.

We concentrate on quantum quenches in two models possessing a free fermionic description.
The first model we considered is the fermionic Anderson model with disorder in the local potential:
\begin{equation}
\hat{H}_{\mathrm{A}} \equiv -t \sum_{j=1}^L \left(\hat{c}_{j}^\dagger \hat{c}_{j+1} + \mathrm{h.c.} \right)
+ \sum_{j=1}^L h_j \hat{c}^\dagger_j \hat{c}_j \;,
\end{equation}
where $\op{c}_{j}^\dagger$ ($\op{c}_{j}$) creates (destroys) a fermion at site $j$,
$t$ is the nearest-neighbor hopping integral and $h_j$ is an uncorrelated on-site random
potential uniformly distributed in the range $\left[ -W/2, W/2 \right]$.
We assume periodic boundary conditions.
It has been mathematically proven~\cite{art:Gertsenshtein_TP_1959}
that, for Hamiltonians like $\hat{H}_{\mathrm{A}}$  and in presence of any $W>0$, 
all the single-particle eigenstates of $\hat{H}_\mathrm{A}$ are exponentially localized.
The second model we considered still describes spinless fermions hopping on a chain, but now
the hopping is long-ranged~\cite{art:Mirlin}:
\begin{equation}\label{eq:fermham}
\op{H}_{\mathrm{lrh}} = \sum_{j_1j_2} t_{j_1 j_2} (\op{c}_{j_1}^\dagger \op{c}_{j_2} + \mathrm{h.c.}) \;,
\end{equation}
where $t_{j_1j_2}$ is a (real) hopping integral between sites $j_1$ and $j_2$. 
We will take the $t_{j_1 j_2}$'s to be random and long-ranged, with a 
Gaussian distribution of zero mean, $\langle t_{j_1j_2} \rangle = 0$, and variance given by:
\begin{equation}
\langle t_{j_1j_2}^2 \rangle = \frac{1}{1+\left( \frac{|j_1-j_2|}{\beta} \right)^{2\gamma}} \,.
\end{equation}
Here $\gamma$ is a real positive parameter setting how fast the hoppings' variance decays with distance.
Notice that, for $j_1=j_2$, we have $\langle t_{j_1j_2}^2 \rangle = 1$ for any $\gamma$, hence
the model has also on-site Gaussian disorder; 
by increasing the distance between the two sites $|j_1-j_2|$,
the variance of the hopping integral decreases with a power law.
The peculiarity of this long-range-hopping model is that, although being one-dimensional and
regardless of the value of $\beta$ (which hereafter is fixed to $1$),
it has an Anderson transition from (metallic) extended eigenstates, for $\gamma<1$, to (insulating) power-law localized eigenstates for 
$\gamma>1$ \cite{art:Mirlin,art:Cuevas,art:Varga}.
Physically, this is due to the fact that, for small $\gamma$, long-range hoppings are capable of overcoming the localization 
due to disorder. 
Having access, in the same model, to physical situations in which the final eigenstates are extended
($\gamma<1$) or localized ($\gamma>1$) will clearly show the role that spatial localization plays in disrupting the 
ability of the GGE to describe the after-quench dynamics. Physically, spatial localization prevents the different ``modes'' of
the system from having an infinite reservoir.

For the considered quenches, we use as initial Hamiltonian $\op{H}_0$
the clean chain with $W=0$ and the same boundary conditions of the final Hamiltonian,
i.e., periodic boundary conditions when quenching to $\HA$ and open
boundary conditions when quenching to $\Hlrh$.
The corresponding initial state $|\Psi_0\rangle$ will be the filled Fermi sea, i.e., the ground state of  
$\op{H}_0$ with $N_{\mathrm{F}}=L/2$, where $N_{\mathrm{F}}$ is the number of fermions.
The reason behind this simple choice for $\op{H}_0$ is that the ``stationary state'' reached does not depend, qualitatively,
on the initial Hamiltonian being ordered or not, see Ref.~\cite{art:Ziraldo_PRL_2012}. 
The final Hamiltonian will be the Anderson model $\HA$ with $W=2$, or the long-range hopping chain
$\Hlrh$ with $\gamma= 0.5$ or $2$.
In all cases the particle number is a constant of motion, therefore
$N_{\mathrm{F}}=L/2$ for any time $t>0$.
To get a smoother size dependence of the computed quantities, the smaller size realizations are
obtained by cutting an equal amount of sites at the two edges of the largest realization.

The two Hamiltonians, being quadratic in the fermion operators, can be diagonalized for any chain of size $L$
in terms of new fermionic operators 
\begin{equation} \label{cmu-cj:eqn}
 \op{d}_\mu^\dagger =  \sum_{j=1}^L u_{j\mu} \op{c}_j^\dagger \;,
\end{equation}
where $u_{j\mu}$ are the wave functions of the eigenmodes of energy $\epsilon_\mu$:
$\op{H}_{\mathrm{A}/\mathrm{lrh}} = \sum_{\mu} \epsilon_\mu \op{d}_\mu^\dagger \op{d}_\mu $. 
The energies $\epsilon_\mu$ and the associated wave functions  $u_{j\mu}$ are obtained, for any
given disorder realization of a chain of size $L$ by numerically diagonalizing the $L \times L$ one-body hopping matrix.

Given an observable $\hat{A}$, consider the two distributions introduced before:
\begin{eqnarray}
\rho_{\D}(A) & \equiv & \sum_{\alpha} |c_\alpha|^2 \delta ( A - A_{\alpha\alpha})   \\
\rho_{\GGE}(A) & \equiv & \sum_{\alpha} \frac{\ee^{-\sum_\mu \lambda_\mu n_\mu^\alpha}}{Z_{\GGE}} 
\delta ( A - A_{\alpha\alpha}) \;,
\end{eqnarray}
where $\delta (x)$ is the Dirac's delta, $\{ \ket{\alpha} \}$ are the many-body eigenstates of $\hat{H}$, 
$A_{\alpha \beta} \equiv \langle \alpha | \hat{A} | \beta \rangle$, 
$c_\alpha \equiv \bra{\alpha} \Psi_0 \rangle$, and $n_\mu^\alpha=0, 1$ is the occupation of 
the single-particle eigenstate $\mu$ in the many-body eigenstate $\ket{\alpha}$.
These functions give the weighted distributions of $A_{\alpha\alpha}$ in the diagonal and GGE ensembles.

Let us discuss a few technical details on the implementation we made, before discussing the physics
emerging from our calculations.
Notice that the sum over $\alpha$ is effectively restricted to the canonical Hilbert space
${\mathcal{H}}_N$ with a fixed number of particles $N=N_{\mathrm{F}}$ in the diagonal ensemble, since
$c_\alpha \equiv \bra{\alpha} \Psi_0 \rangle = 0$ if $N_{\alpha}\neq N_{\mathrm{F}}$.
No such restriction is in principle present in the GGE case, where the sum over $\alpha$ runs
over the grand-canonical Hilbert space.
By definition, the distributions are such that $\dav{\hat{A}} = \int \! A \, \rho_{\D}(A) \, \ud A$ 
and $\ggeav{\hat{A}} = \int \ud A \, \rho_{\GGE}(A) A$, 
where the integration is over the domain of $A_{\alpha\alpha}$.
As customary in any numerical finite-size study, one really needs to consider a coarse-grained
version of these distributions, obtained by splitting the domain of $A$ into small intervals $\Delta^{(i)}$
of amplitude $\Delta$. 
Such a coarse-grained distribution has exactly the form of a weighted density of states,
see Eq.~\eqref{eq:wdist}, with
$w_\alpha = \left| c_\alpha \right|^2 / \Delta$ in the diagonal case, and 
$w_\alpha = \ee^{-\sum_\mu \lambda_\mu n_\mu^\alpha}/(\Delta \; Z_{\GGE})$
in the GGE case.
The configuration space $\{\ket{\alpha}\}$ (i.e., the canonical Hilbert space
${\mathcal{H}}_N$ for the diagonal distribution and the full Hilbert space
for the GGE) over which the two weighted distributions are defined is discrete and grows exponentially with the system size. 
The weighted-WLA is therefore 
the appropriate tool for the numerical computation of $\rho_{\D}(A)$ and $\rho_{\GGE}(A)$. 
The eigenstates $\ket{\alpha}$ which appear in the definition of $\rho_{\D}(A)$ 
have a fixed number of fermions $N_{\mathrm{F}}$ (the same of the initial state),
while in $\rho_{\GGE}(A)$ the number of particles can change.
In the weighted-WLA, for the diagonal ensemble, we use therefore a ``particle conserving'' proposal scheme: 
given a state $\ket{\alpha}$, the state $\ket{\alpha^\prime}$ is given by moving at random a fermion 
in one of the unoccupied single-particle eigenstates.
In this case, the ratio $w_{\alpha^\prime}/w_{\alpha}$ which appears in Eq.~\eqref{eq:MCstepw}, is equal to:
\[
\frac{w_{\alpha^\prime}}{w_{\alpha}} = \frac{\left| c_{\alpha^\prime} \right|^2}{\left| c_\alpha \right|^2} \;,
\]
where the coefficient $|c_\alpha|^2$ is the square of the determinant of a $N_{\mathrm{F}} \times N_{\mathrm{F}}$ 
matrix (see \cite[App.~D]{Ziraldo:thesis} for the explicit expression of $|c_\alpha|^2$).
For the GGE case, instead, we do not have restrictions on the number of fermions and, 
given a state $\ket{\alpha}$, we generate a state $\ket{\alpha^\prime}$ by changing 
the occupation of a randomly selected single-particle eigenstate $\mu$.
In this case:
\[
\frac{w_{\alpha^\prime}}{w_{\alpha}} = \ee^{ \pm \lambda_\mu} \;,
\]
where the $+$ ($-$) sign appears when the mode $\mu$ is initially occupied (empty).
Let us recall that the Lagrange's multipliers $\lambda_\mu$ are obtained by requiring
$\bra{\Psi_0} \hat{d}_\mu^\dagger \hat{d}_\mu  \ket{\Psi_0}  = \ggeav{\hat{d}_\mu^\dagger \hat{d}_\mu }$.
This condition, written explicitly, reads:
\begin{equation} \label{eq:GGEoccExplicit}
\ee^{\lambda_\mu} = \frac{1}{\sum_\nu n_\nu^0 \left| \left[ u^{0\dagger} u \right]_{\nu \mu} \right|^2} - 1 \;,
\end{equation}
where $u^0$ and $u$ are $L \times L$ matrices whose elements $u_{j\nu}^0$ and $u_{j\mu}$
are the single-particle wavefunctions of the initial Hamiltonian $\hat{H}_0$ and the final one $\hat{H}$, 
and $n^0_\nu = 0,1$ is the occupation of the $\nu$th eigenstate of $\hat{H}_0$ in the initial state.
The difference in the computational effort on computing the ratio $w_{\alpha^\prime}/w_{\alpha}$
in the two ensembles is evident: in the diagonal case at each step we have to compute the determinant of a 
$N_{\mathrm{F}} \times N_{\mathrm{F}}$ matrix, while in the GGE we have just to recover
the value of $\ee^{ \lambda_\mu}$ (they can be computed and stored before the
Monte Carlo calculation because their number is $L$). Here we will show results for sizes up to $L=256$,
where both $\rho_{\D}(A)$ and $\rho_{\GGE}(A)$ can be computed and compared.
(For the GGE ensemble, we could reach $L=1024$ without problem.)
In the numerical computations we used a minimum value of the WL parameter
$\epsilon=\ln f_{\rm min}=10^{-6}$, and we split the domain of $A$ in $L$ bins. 
Notice that the domain of $A$ in $\rho_{\GGE}(A)$ is always larger than
the domain of $\rho_{\D}(A)$ because, in the GGE, the many-body eigenstates
do not have a restriction on the number $N_\mathrm{F}$ of fermions.

In the next two subsections we show the results obtained with the weighted-WLA
for the calculation of $\rho_{\D}(A)$ and $\rho_{\GGE}(A)$ for two observables,
the total energy and the local density.
The physical picture emerging from the calculation of the full distribution function of 
the after-quench energy and local-density confirms and extends the results discussed in 
Ref.~\cite{art:Ziraldo_PRL_2012,art:Ziraldo_PRB_2013}. In particular, we find clear differences between the
diagonal and GGE distributions, even at the level of the variances, whenever a disorder-induced 
spatial localization is at play in the after-quench Hamiltonian.

\subsection{Probability distributions of the energy}
\begin{figure}[ht]
\begin{center}
	\includegraphics[width=0.85\textwidth]{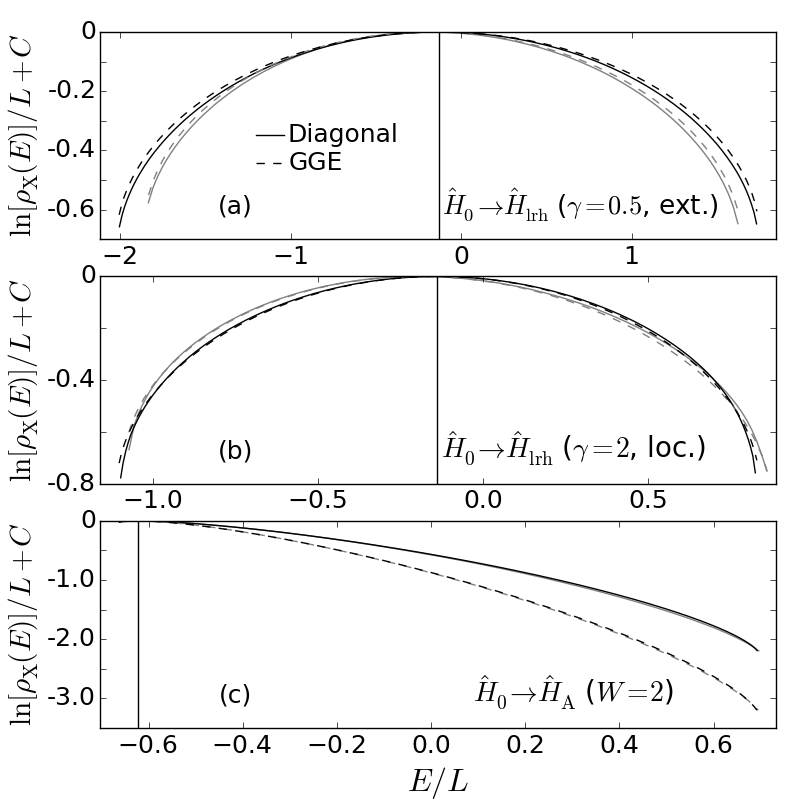}
\end{center}
\caption{Value of $\ln[ \rho_{\D}(E) ]/L$ and $\ln[ \rho_{\GGE}(E)]/L$ 
computed with the weighted-WLA. 
The gray curves are obtained with $L=128$, while the black ones with $L=256$.
The solid vertical lines are the average energy after the quench, i.e. 
$\bra{\Psi_0} \hat{H} \ket{\Psi_0}$, for $L=256$.
The three panels are obtained starting from the ground states of clean chains and 
quenching to different disordered Hamiltonians:
panel (a) long-range hopping with $\gamma=0.5$ (extended eigenstates), 
panel (b) long-range hopping with $\gamma=2$ (localized eigenstates) and
panel (c) Anderson model with $W=2$.
For the computations we used a single disorder realization
and, to get a smoother size dependence, the smaller size realization is obtained by cutting 
an equal amount of sites at the two edges of the larger realization. These distributions are
obtained for a single realization of the couplings, but we verified that, for large sizes, the results are self-averaging.
}
\label{fig:rho-GGE-D}
\end{figure}

\begin{figure}[tbh]
\begin{center}
       \includegraphics[width=0.85\textwidth]{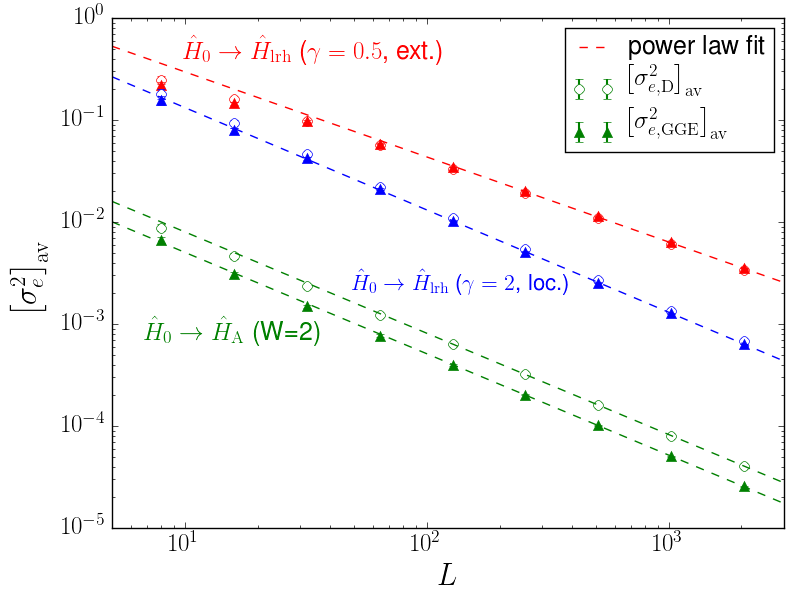}
\end{center}
\caption{Variances $\sigma^2_{e,\D}=\sigma^2_{E,\D}/L^2$ (empty circles) and 
$\sigma^2_{e,\GGE}= \sigma^2_{E,\GGE}/L^2$ (solid triangles) as a function of the size $L$. 
The data are obtained using the same set of quenches used in Fig.~\ref{fig:rho-GGE-D} and 
the values are computed using Eqs.~\eqref{eq:sigmaE-GGE-D}.
Error bars are calculated by averaging over 20 different realizations of the disorder.
The dashed lines are power law fits $\sigma^2_{e}\sim L^{-s}$, where $s \approx 1$ for the Anderson case,
while, for the quench to $\Hlrh$, $s \approx
0.82$ when $\gamma = 0.5$, and $s \approx 0.95$ when $\gamma = 2$.
Notice the observable difference between $\sigma^2_{e,\D}$ and $\sigma^2_{e,\GGE}$ when the final eigenstates
are localized.}
\label{fig:sigmaE-GGE-D}
\end{figure}

%
The first observable we consider is the total energy: 
Here $A_{\alpha\alpha} \to E_{\alpha} = \sum_{\mu} \epsilon_\mu n_\mu^\alpha$,
where $n_\mu^\alpha= \bra{\alpha} \hat{d}_\mu^\dagger \hat{d}_\mu \ket{\alpha} = 0,1$ 
are the single-particle occupations of the eigenstate $\ket{\alpha}$.
In Fig.~\ref{fig:rho-GGE-D} we show 
the distributions $\ln[\rho_{\D}(E)]/L$ and $\ln[\rho_{\GGE}(E)]/L$, computed for $L=128$ and $L=256$, 
for the three cases we have studied, i.e., quenches from an initially ordered half-filled chain $\op{H}_0$ towards: 
1) a long-range hopping Hamiltonian $\Hlrh$ with extended eigenstates ($\gamma=0.5$, top), 
2) $\Hlrh$ with localized eigenstates ($\gamma=2$, center),
and 3) an Anderson Hamiltonian $\HA$ with a disorder width $W=2$ (bottom).
%
%
Observe, first, that the distributions $\rho_{\D}(E)$ and $\rho_{\GGE}(E)$ shown in Fig.~\ref{fig:rho-GGE-D}
have identical average (denoted by a solid vertical line)
\[
\dav{\hat{H}} = \int \! \ud E \, \rho_{\D}(E) \, E = \int \! \ud E \,  \rho_{\GGE}(E) \, E = \ggeav{\hat{H}} \;. 
\]
This result comes directly from the fact that the energy does not fluctuate in time
(i.e., the diagonal energy coincides with the average energy $\bra{\Psi_0} \hat{H}\ket{\Psi_0}$)
and GGE fixes the occupation of the fermionic eigenstates in such a way as to exactly
reproduce $\bra{\Psi_0} \hat{H}\ket{\Psi_0}$.
The form of the two distributions, however, differs considerably, most notably at the extremes of the spectrum, and
for the Anderson model case.
Let us now consider the fluctuations of the energies in both distributions. 
In the diagonal ensemble the variance is:
\begin{equation} \label{sigma2:eqn}
\sigma^2_{E,\D} = \int \ud E \, \rho_{\D}(E) \, E^2 - \dav{\hat{H}}^2 = 
\dav{\hat{H}^2} - \dav{\hat{H}}^2  \;,
\end{equation}
where the expression on the right-hand side holds only for the Hamiltonian 
(it would not apply to arbitrary operators, because $(A_{\alpha\alpha})^2\neq \bra{\alpha}\hat{A}^2\ket{\alpha}$).
An entirely similar expression applies to the GGE case.
Since the energy is an extensive operator, it is reasonable to ask what happens to the
fluctuations in the energy-per-site $e=E/L$, which are simply given by 
$\sigma^2_{e,\D}=\sigma^2_{E,\D}/L^2$, and $\sigma^2_{e,\GGE}=\sigma^2_{E,\GGE}/L^2$. 
On pretty general grounds, for quenches of local non-integrable Hamiltonians, 
it is known~\cite{art:Rigol_Nat_2008,art:BiroliPRL} that $\sigma^2_{e,\D}\to 0$ in
the thermodynamic limit, $L\to \infty$.
Indeed, as shown in Fig.~\ref{fig:sigmaE-GGE-D} both $\sigma_{e,\D}^2$ and $\sigma_{e,\GGE}^2$
decrease to $0$ for $L\to \infty$ for the three considered cases. 
For our quadratic problems, however, we can say a bit more. 
First of all, from the explicit expression in Eq.~\eqref{sigma2:eqn} after
very simple algebra (mainly using Wick's theorem), we arrive at:
\begin{eqnarray}
\sigma_{e,\GGE}^2 = \frac{1}{L^2}\sum_\mu \epsilon_\mu^2 n^0_\mu \left( 1 - n^0_\mu \right) \;, \\
\sigma_{e,\D}^2 = \sigma_{e,\GGE}^2
- \frac{1}{L^2} \sum_{\mu_1 \neq \mu_2 } \epsilon_{\mu_1} \epsilon_{\mu_2} \left| G^0_{\mu_1 \mu_2} \right|^2 \;,
\label{eq:sigmaE-GGE-D}
\end{eqnarray}
where $G^0_{\mu_1 \mu_2} \equiv \bra{\Psi_0} \hat{d}^\dagger_{\mu_1} \hat{d}_{\mu_2} \ket{\Psi_0}$ 
is the $t=0$ one-body Green's function.
The off-diagonal elements of $G^0_{\mu_1 \mu_2}$ play here an important role, and the second term in $\sigma_{e,\D}^2$
originates from the fact that, by definition, GGE does not include correlations between different
eigen-modes, i.e., $\ggeav{\hat{d}^\dagger_{\mu_1} \hat{d}_{\mu_2}}=0$, when $\mu_1 \neq \mu_2$.
%

Let us first consider the Anderson model case. Assuming, as done so far, a bounded
distribution of disorder, we are guaranteed that a finite bound $\epsilon_{\mathrm{max}}$ 
exists such that $|\epsilon_\mu| \leq \epsilon_{\mathrm{max}}$ for any $L$. 
With this assumption, it is easy show that $\sigma_{e,\GGE}^2$ has to go to zero at least 
as $1/L$ for $L\to \infty$.
Indeed, the occupation factors appearing in $\sigma_{e,\GGE}^2$ are such that
$0\le n^0_\mu \left( 1 - n^0_\mu \right) \le 1/4$. Hence: 
\begin{equation}\label{eq:ggebound}
\sigma_{e,\GGE}^2 \leq \frac{\epsilon_{\mathrm{max}}^2}{L^2}\sum_\mu n^0_\mu \left( 1 - n^0_\mu \right)
\leq \frac{\epsilon_{\mathrm{max}}^2}{4L} \;.
\end{equation}
%
The same statement can be made for $\sigma_{e,\D}^2$, because the difference
between the two variances has a similar upper bound:
\begin{equation}\label{eq:dbound}
\hspace{-20mm} |\sigma_{e,\D}^2 - \sigma_{e,\GGE}^2 | \leq
\frac{1}{L^2} \sum_{\mu_1 \neq \mu_2 } |\epsilon_{\mu_1}| |\epsilon_{\mu_2}| \left| G^0_{\mu_1 \mu_2} \right|^2 
\leq \frac{\epsilon_{\mathrm{max}}^2}{L^2} \sum_{\mu_1 \neq \mu_2 } \left| G^0_{\mu_1 \mu_2} \right|^2 
\leq \frac{\epsilon_{\mathrm{max}}^2}{L} \;,
\end{equation}
where we used that $\sum_{\mu_1\mu_2} \left| G^0_{\mu_1 \mu_2} \right|^2 = N_{\mathrm{F}} \leq L$.
%
Nevertheless, although both $\sigma_{e,\D}^2$ and $\sigma_{e,\GGE}^2$ go to $0$ as $1/L$ for the Anderson model, 
they do so with a different pre-factor, see Fig.~\ref{fig:sigmaE-GGE-D} and comments below. 

For the quenches to $\Hlrh$, a bound $\epsilon_{\mathrm{max}}$ for the single-particle 
spectrum is in principle not defined: one can think of rare realizations in which the 
hopping is large at arbitrarily large distances, which would
give an unbounded distribution of eigenvalues $\epsilon_{\mu}$. 
Indeed, the behavior of both $\sigma_{e,\D}^2$ and $\sigma_{e,\GGE}^2$ suggest, 
see Fig.~\ref{fig:sigmaE-GGE-D}, that the power-law approach to $0$ might be slower
than $1/L$, i.e., as $L^{-s}$ with $s<1$ (we find $s \approx 0.82$ for the case $\gamma=0.5$
and $s \approx 0.95$ for $\gamma=2$). 
While this might be a finite-size artifact, we find it intriguing that such deviations are quite clearly seen 
for quenches to $\Hlrh$: they might be due to the power-law nature of the hopping integral variance.
 
\begin{figure}[tbh]
\begin{center}
		\includegraphics[width=0.85\textwidth]{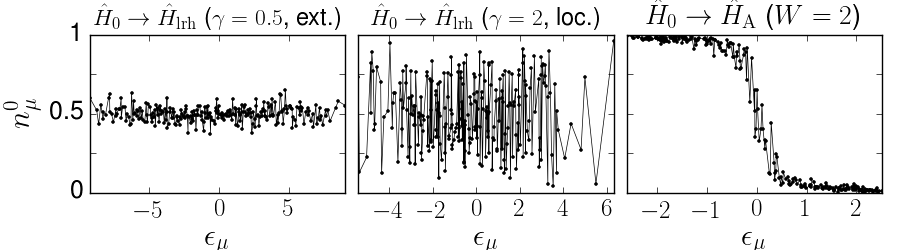}  \\
Panel (a) \\
		\includegraphics[width=0.85\textwidth]{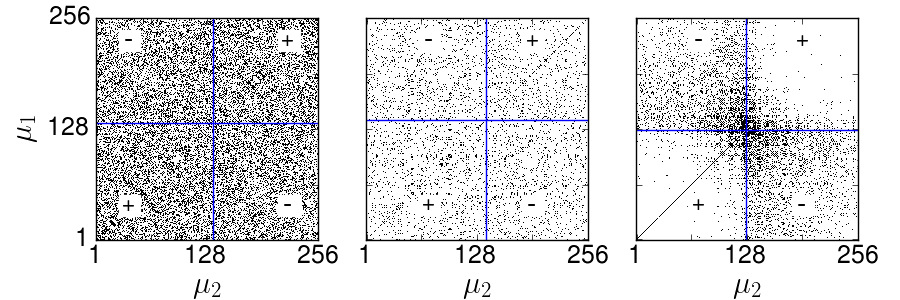} \\
Panel (b) \\
\end{center}
\caption{
Panel (a): occupations $n_{\mu}^0 = \bra{\Psi_0}\hat{d}_\mu^\dagger \hat{d}_\mu \ket{\Psi_0}$ 
as a function of the single-particle energy $\epsilon_\mu$. 
Panel (b): representation of the matrix $| G^0_{\mu_1 \mu_2}|^2$.
For the diagonal and off-diagonal elements we add a black pixel when the value
exceeds their mean value. For the diagonal elements the mean value is $x \equiv \sum_\mu (n_{\mu}^0)^2/L$,
while for the off-diagonal elements the mean value is $(N_\mathrm{F} - x L )/L(L-1)$, where $N_\mathrm{F}$
is the number of fermions in the initial state, and we used the relation 
$\sum_{\mu_1\mu_2} | G^0_{\mu_1 \mu_2} |^2 = N_\mathrm{F}$ (see \cite[App.~D]{Ziraldo:thesis}).
The vertical and horizontal lines indicate the indexes at which the single-particle energies $\epsilon_{\mu_1}$
and $\epsilon_{\mu_2}$ change sign, and the signs shown in the four quadrants
are those of the product $\epsilon_{\mu_1} \epsilon_{\mu_2}$.
For the two panels we used $L=256$ and the same quenches used in Fig.~\ref{fig:rho-GGE-D}
and Fig.~\ref{fig:sigmaE-GGE-D}.}
\label{fig:occAndMatrices}
\end{figure}

Concerning the similarity between $\sigma_{e,\D}^2$ and $\sigma_{e,\GGE}^2$,
we observe that the two essentially coincide for the case of a quench to $\Hlrh$ with extended eigenstates, 
while there is a small discrepancy for the quench to $\Hlrh$ with localized eigenstates, and a
quite clear different pre-factor in the Anderson model case,
$\sigma_{e,\D}^2\sim C_{\D}/L$ and $\sigma_{e,\GGE}^2\sim C_{\GGE}/L$ with $C_{\GGE}<C_{\D}$.
This different pre-factor can be understood by analyzing the term
$\sum_{\mu_1 \neq \mu_2 } \epsilon_{\mu_1} \epsilon_{\mu_2} | G^0_{\mu_1 \mu_2} |^2$
which appears in Eq.~\eqref{eq:sigmaE-GGE-D}.
In Fig.~\ref{fig:occAndMatrices}, panel (b), we show the structure of the matrix $| G^0_{\mu_1 \mu_2} |^2$
for the three quench cases.
We divide this matrix into four sectors, one for each sign of the single-particle energies
$\epsilon_{\mu_1}$ and $\epsilon_{\mu_2}$: in two of these quadrants the product
$\epsilon_{\mu_1}\epsilon_{\mu_2}$ is positive (top-right and bottom-left), and
in the others is negative.
For quenches to $\Hlrh$ this matrix is almost equally distributed in all the four sectors: 
the sum $\sum_{\mu_1 \neq \mu_2 } \epsilon_{\mu_1} \epsilon_{\mu_2} | G^0_{\mu_1 \mu_2} |^2$
has cancellations, leading to $\sigma_{e,\GGE}^2 \approx \sigma_{e,\D}^2$
for large sizes.
For quenches to $\HA$, on the contrary, the matrix $| G^0_{\mu_1 \mu_2} |^2$ 
is mainly concentrated in the sectors in which 
$\epsilon_{\mu_1} \epsilon_{\mu_2} < 0$, leading to $\sigma_{e,\GGE}^2 < \sigma_{e,\D}^2$.
%
%
%

Finally, let us comment on one aspect of the distributions shown in Fig.~\ref{fig:rho-GGE-D} which can be easily 
understood from the single-particle occupations shown in Fig.~\ref{fig:occAndMatrices}.
We see that, when the after-quench Hamiltonian is the Anderson model, 
$\rho_\D (E)$ has both mode (i.e., maximum value) and average very close 
to the ground state energy: the quench excites mostly the low-energy part of the many-body spectrum.
On the contrary, for both the quenches towards $\Hlrh$, mode and average
are almost in the middle of the many-body spectrum; there, indeed, the quench is more dramatic: 
we are going from the ground state of a chain with nearest-neighbor hopping to a disordered chain with 
long-range hopping. This is evident by looking at the occupations
$n_\mu^0 \equiv \bra{\Psi_0} \hat{d}^\dagger_\mu \hat{d}_\mu \ket{\Psi_0}$ as a function 
of the single-particle energy $\epsilon_\mu$, shown in Fig.~\ref{fig:occAndMatrices}, panel (a).
By definition, only the eigenstates of $\hat{H}_0$ with $\epsilon_\nu^0<0$ are occupied in $\ket{\Psi_0}$.
The quench to $\HA$ only slightly modifies the initial occupations:
$n_\mu^0$, apart for fluctuations due to disorder, goes smoothly from $1$, in the lower 
part of the single-particle spectrum, to $0$, in the highest part of the 
spectrum.
On the contrary, for the quenches towards $\Hlrh$, the single-particle spectrum is entirely excited, both the 
positive and the negative energy part.
This explains why, for these quenches, the after-quench energy $\bra{\Psi_0} \hat{H} \ket{\Psi_0}
= \sum_\mu \epsilon_\mu n^0_\mu$ is near the center of the many-body spectrum.

\subsection{Probability distributions of the local density}
%
Let us now consider the local density $\hat{n}_j \equiv \hat{c}^\dagger_j \hat{c}_j$,
perhaps the simplest one-body observable.
For definiteness, we concentrate on $j=L/2$, the center of the chain.
It is important to stress that we are going to consider the fluctuations of   
$\hat{n}_j$ {\em before} any possible average over the sites $j$: averaging over the sites
$j$ an intensive local operator would effectively send to zero the fluctuations
in the thermodynamic limit \cite{art:BiroliPRL}, while we will show that, for
a fixed $j$, finite fluctuations survive in the thermodynamic limit when the eigenstates are 
localized, due to disorder.
 
\begin{figure}[tbh]
\begin{center}
	\includegraphics[width=0.85\textwidth]{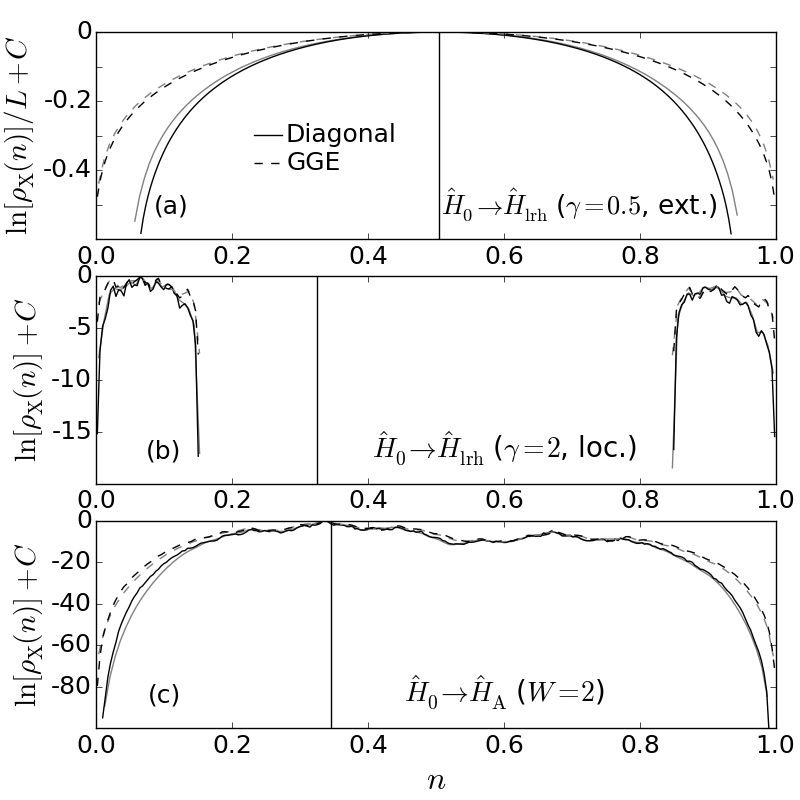}
\end{center}
\caption{
Distributions of the local density $\hat{n}_{j}$ at the center of the chain, $j=L/2$, for the
diagonal ensemble and the GGE, and for $L=256$ (black curves) or $L=128$ (gray curves).
In panel (a), we plot $\ln[ \rho_{\D}(n) ]/L$  and $\ln[ \rho_{\GGE}(n)]/L$,
while in panels (b) and (c) $\ln[ \rho_{\D}(n) ]$  and $\ln[ \rho_{\GGE}(n)]$.
The vertical lines are the diagonal and GGE average of $\hat{n}_{j}$, which coincide
for the local density.
The three panels are obtained using the same quenches of Fig.~\ref{fig:rho-GGE-D}.}
\label{fig:rho-n-GGE-D}
\end{figure}

\begin{figure}[tbh]
   \begin{center}
	\includegraphics[width=0.85\textwidth]{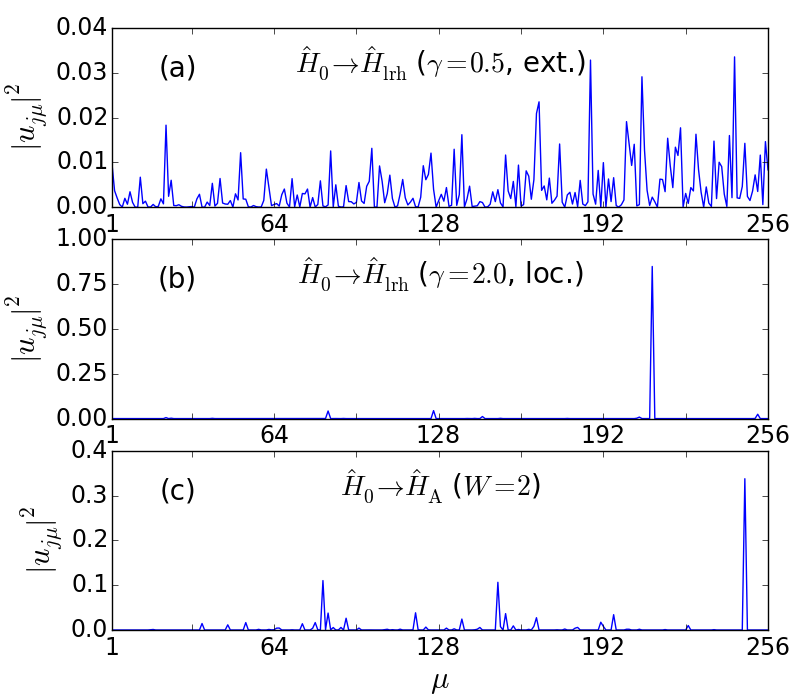}
   \end{center}
\caption{Squared single-particle wavefunction $|u_{j\mu}|^2$ as a function of the
eigenstates index $\mu$, at fixed site $j=L/2$. We have taken $L=256$ and the three panels 
are obtained using the same quenches of Fig.~\ref{fig:rho-GGE-D}.}
\label{fig:wl-local-occ}
\end{figure}

The diagonal and GGE distributions 
$\rho_{\D}(n)$ and $\rho_{\GGE}(n)$ are now constructed using the matrix elements
$n_{\alpha \alpha} \equiv \bra{\alpha} \hat{n}_j \ket{\alpha} = \sum_\mu |u_{j\mu}|^2 n_\mu^\alpha$,
where $n_\mu^\alpha=0,1$ are, as before, the single-particle occupations of the eigenstate $\ket{\alpha}$.
In Fig.~\ref{fig:rho-n-GGE-D} we plot $\ln[ \rho_{\D}(n) ]$ and $\ln[ \rho_{\GGE}(n)]$,
computed for the three quenches discussed before.
%
%
The case of a quench to $\Hlrh$ with $\gamma=2.0$ (localized eigenstates) is quite peculiar. 
The values that $n$ can assume is actually split in two separated domains, one just above 
$n=0$ and one just below $n=1$, and the mean value is exactly in the middle,
where no values of $n_{\alpha\alpha}$ happen to fall.
This is due to the strong spatial localization of the eigenstates.
As we show in Fig.~\ref{fig:wl-local-occ}, at fixed $j$, the value of $|u_{j\mu}|^2$
is strongly localized in a single eigenstate $\tilde{\mu}$. This implies that the value
$n_{\alpha \alpha} = \sum_\mu |u_{j\mu}|^2 n_\mu^\alpha$ has a strong jump when
we move from a state $\ket{\alpha}$ in which $n_{\tilde{\mu}}^\alpha=0$,
to the state $\ket{\alpha}$ in which $n_{\tilde{\mu}}^\alpha=1$.
For the quench to $\HA$, with $W=2$, the localization is not strong enough to produce
such a gap: we however expect this to happen for larger values of the disorder amplitude $W$.
%
%
%

\begin{figure}[tbh]
\begin{center}
	\includegraphics[width=0.85\textwidth]{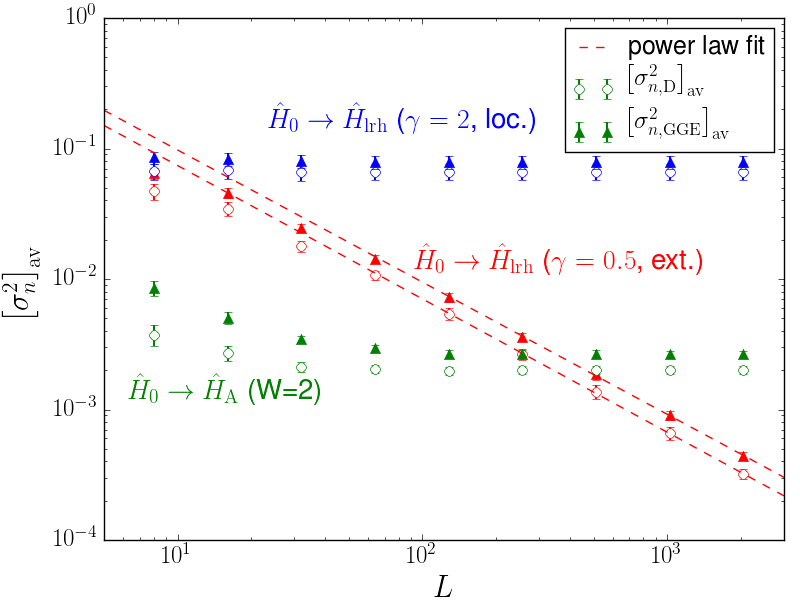}
\end{center}
\caption{Plot of the variances $\sigma_{n,\D}^2$ (circles) and $\sigma_{n,\GGE}^2$ (triangles) 
as a function of size. The data are obtained using the same set of 
quenches used in Fig.~\ref{fig:sigmaE-GGE-D} and
the values are computed using Eq.~\eqref{eq:sigmaLocalDensity-GGE-D}.
Error bars are calculated by averaging over 20 different realizations of the disorder.
The dashed lines are power-law fits $\sigma^2_n \sim L^{-s}$, where
$s \approx 1$ in both cases.}
\label{fig:sigmaLocalDensity-GGE-D-new}
\end{figure}

Since $\hat{n}_j$ is a one-body operator, the diagonal and GGE averages coincide
\cite{art:Ziraldo_PRB_2013}, and therefore, the mean value of the two distributions is the same:
\[
\int \! \ud n \, \rho_{\D}(n) \, n = \int \! \ud n \, \rho_{\GGE}(n) \, n \; .
\]
We also note that, for this observable, $\hat{n}_j^m = \hat{n}_j$ for any positive
integer $m$, and therefore $\dav{\hat{n}_j^m} = \ggeav{\hat{n}_j^m}$.
However, this does not allow us to conclude that the two distributions 
$\rho_{\D}(n)$ and $\rho_{\GGE}(n)$ coincide, since, unlike the case of the total energy, we have that,
for instance:
\[
\int \! \ud n \, \rho_{\D/\GGE}(n) \, n^2  \neq \langle \hat{n}_j^2\rangle_{\D/\GGE}  \;.
\]
The variance of the two distributions can be computed by exploiting again Wick's theorem. 
We find that:
\begin{eqnarray}
\sigma_{n,\GGE}^2 = \sum_\mu \left| u_{j\mu} \right|^4 n^0_\mu \left( 1 - n^0_\mu \right) \\
\sigma_{n,\D}^2 = \sigma_{n,\GGE}^2
 - \sum_{\mu_1 \neq \mu_2 } \left| u_{j\mu_1} \right|^2 \left| u_{j\mu_2} \right|^2 
    \left| G^0_{\mu_1 \mu_2} \right|^2 \;.
\label{eq:sigmaLocalDensity-GGE-D}
\end{eqnarray}
In Fig.~\ref{fig:sigmaLocalDensity-GGE-D-new} we plot
$\sigma_{n,\GGE}^2$ and $\sigma_{n,\D}^2$ as a function of size. 
We see that, in both ensembles, the variances vanish as $1/L$ when quenching to $\Hlrh$ with $\gamma = 0.5$
(extended eigenstates) while they are {\em finite} when quenching to $\HA$ and to $\Hlrh$ with $\gamma = 2$,
i.e., when the final Hamiltonian has localized eigenstates.
These results agree with the findings of Ref.~\cite{art:Kai_PRA_2013}, who show that,
for large $L$, the variance of few-body intensive (but not site-averaged) observables 
remains finite both in the microcanonical ensemble and in the diagonal ensemble 
for the Aubry-Andr\'e model. 

From the equation for $\sigma_{n,\GGE}^2$, we see that it is related to an inverse participation
ratio (IPR): the sum is over the eigenstates $\mu$,
%
%
each $\mu$ weighted with the corresponding occupation factor 
$0\le n^0_\mu \left( 1 - n^0_\mu \right)\le 1/4$ depending on the initial state. 
It is therefore easy to realize that: 
\begin{equation}
\sigma_{n,\GGE}^2 \leq \frac{1}{4} \sum_\mu \left| u_{j\mu} \right|^4 = \frac{\IPR_{j}}{4} \;,
\end{equation}
where the last equality defines the IPR at fixed site $j$.
This shows that, whenever the IPR goes to zero, i.e., when the final Hamiltonian has delocalized 
eigenstates, $\sigma_{n,\GGE}^2$ goes to zero as well.
For a final Hamiltonian with localized eigenstates we have instead the opposite: there is at least
one eigenstate $\tilde{\mu}$ localized around $j$, and therefore there is a single-particle wavefunction 
$u_{j \tilde{\mu}}$ which does not vanish in the thermodynamic limit; 
if the initial occupation $n^0_{\tilde{\mu}}$ of this eigenstate is such that $0<n^0_{\tilde{\mu}}<1$,
then $\sigma_{n,\GGE}^2$ remain finite in the thermodynamic limit.
 
Concerning $\sigma_{n,\D}^2$, Eq.~\eqref{eq:sigmaLocalDensity-GGE-D} can be rewritten as:
\begin{equation}
\sigma_{n,\D}^2 = \sigma_{n,\GGE}^2 -\delta^2_{jj} \;, 
\end{equation}
where $\delta^2_{jj}$ denotes the mean squared time-fluctuations of the single-particle Green's function
$G_{j_1j_2}(t)$~\cite{art:Ziraldo_PRL_2012}:
\begin{equation} \label{eq:defDelta2}
\delta_{j_1j_2}^2 \equiv \lim_{t \rightarrow \infty}\frac{1}{t} \int_0^t \ud t^\prime \,
\left| \delta G_{j_1j_2}(t^\prime) \right|^2 \;,
\end{equation}
$\delta G_{j_1j_2}(t) \equiv G_{j_1j_2}(t) - \tav{G_{j_1j_2}}$ being the time
fluctuation with respect to the long-time average $\tav{G_{j_1j_2}}$.
Physically, $\delta_{jj}^2$ is the averaged long-time fluctuation of the local density $\hat{n}_j=\hat{c}_j^\dagger \hat{c}_j$.
In Refs.~\cite{art:Ziraldo_PRL_2012,art:Ziraldo_PRB_2013} we have shown that if the final Hamiltonian
has extended eigenstates, then $\delta^2_{jj} \approx 1/L$ for large sizes, while
$\delta^2_{jj}$ remains finite when the final Hamiltonian has localized eigenstates. 
This explains all the features shown in Fig.~\ref{fig:sigmaLocalDensity-GGE-D-new},
in particular the clear difference between $\sigma_{n,\D}^2$ and $\sigma_{n,\GGE}^2$ in all cases. 
%

\section{Summary and conclusions}\label{sec:conclusions}

In this paper we have introduced a Monte Carlo method --- obtained by a rather natural 
extension of the Wang-Landau algorithm~\cite{art:WL-PRL,art:WL-PRE,art:WL-CC} ---
which allows to compute quite general weighted distribution functions of the form relevant 
to quantum quenches, see  Eq.~\eqref{eq:wdist-intro}.
We have used this approach to analyze quantum quenches for free-fermion Hamiltonians in presence of disorder. 
For these systems, thanks to Wick's theorem, after-quench expectation values and time
averages require a modest computational effort, proportional to a power-law of the size $L$ \cite{art:Ziraldo_PRB_2013}. 
However, the calculation of full probability distributions 
--- like the diagonal ensemble distribution $\rho_{\mathrm{D}}(A)$, Eq.~\eqref{eq:rhod}, or the
GGE one $\rho_{\mathrm{GGE}}(A)$, Eq.~\eqref{eq:rhogge} ---
would still require a sum over an exponential number of terms, hence unfeasible beyond very small sizes.

Although quadratic, hence with an extensive number of conserved quantities, these free-fermion 
problems are not described by the GGE ensemble whenever the disorder is such that the after-quench eigenstates
are localized. More precisely, while the GGE ensemble is known to correctly capture the long-time average of
any one-body operator, almost ``by construction''  \cite{art:Ziraldo_PRB_2013}, it does not capture correlations
induced by the spatial localization of the eigenstates. Our study further explored this issue by explicitly calculating
and comparing the full probability distributions of both the energy and the local density in the two relevant ensembles.

Concerning the energy, we have explicitly verified that the form of the two distributions for the diagonal and GGE ensembles 
differs considerably, most notably at the extremes of the spectrum, and for the Anderson model case. 
More in detail, we have verified that, regardless of the final Hamiltonian, the averaged fluctuations
of the energy-per-site, $\left[ \sigma_{e,\D}^2\right]_{\mathrm{av}}$ and $\left[\sigma_{e,\GGE}^2\right]_{\mathrm{av}}$, go to zero
in the thermodynamic limit, see Fig.~\ref{fig:sigmaE-GGE-D}, in agreement with the general analysis of Refs.~\cite{art:Rigol_Nat_2008,art:BiroliPRL}. 
Nevertheless, we find that there is a clearly detectable difference in the two variances
when the final Hamiltonian has localized eigenstates.

In addition to the energy, we studied the local density distributions.
For this observable, it was already known that, even in presence of disorder and localization,
the GGE expectation value coincides with the diagonal average~\cite{art:Ziraldo_PRB_2013}, 
a property true, more generally, for any one-body operator~\cite{art:Ziraldo_PRB_2013}.
Our numerical results confirm that even if the averages of  $\rho_{\mathrm{D}}(n)$ 
and $\rho_{\mathrm{GGE}}(n)$ coincide, the two distributions are different when localization is present, 
with clearly detectable differences already at the level of the variance, see Fig.~\ref{fig:sigmaLocalDensity-GGE-D-new}:
$\sigma_{n,\GGE}^2$ and $\sigma_{n,\D}^2$ differ by a quantity which represents the averaged long-time fluctuations of the local 
density \cite{art:Ziraldo_PRL_2012,art:Ziraldo_PRB_2013}, which remain finite whenever the final Hamiltonian has localized eigenstates.

Other many-body operators, like density-density correlations, might be analyzed in a similar way.
Here, even the average values are not in general well described by the GGE distribution whenever localization 
is at play \cite{art:Ziraldo_PRB_2013}: we expect, once again, clearly visible discrepancies between the diagonal and GGE distributions
in such cases.

In conclusion, as explained above, the weighted-WLA we have presented circumvents the 
difficulty associated to the exponentially large Hilbert space to be visited, even when the relevant ingredients entering the distribution 
--- matrix elements and overlap between states --- can be calculated quite effectively. 
In principle, the applicability of the method is not limited to ``quadratic fermion problems'' 
of the type we have considered in the paper: 
If, by Bethe Ansatz, or any other exact technique or even by a suitable quantum Monte Carlo approach, 
one would be able to calculate matrix elements and overlaps, the method we have illustrated would provide an effective Monte Carlo
sampling of the relevant distribution functions.

\ack 
We acknowledge discussions with A. Laio, A. Russomanno and A. Silva.
Research was supported by MIUR, through PRIN-2010LLKJBX-001, by SNSF, 
through SINERGIA Project CRSII2 136287\ 1, by the EU-Japan Project LEMSUPER, and 
by the EU FP7 under grant agreement n. 280555.

\section*{References} 

\end{document}